\documentclass[onecolumn]{aa}

\usepackage{txfonts}

\usepackage{graphicx}

\usepackage{natbib}
\bibpunct{(}{)}{;}{a}{}{,}

\begin{document}

\title{VIRGO sensitivity to binary coalescences and the Population III black hole binaries}

\author{
K. Kulczycki\inst{1}
\and
T. Bulik\inst{2,1}
\and
K. Belczy\'nski\inst{3,4}
\and
B. Rudak\inst{1}
}

\institute{
Nicolaus Copernicus Astronomical Center,
Bartycka 18, 00716 Warsaw, Poland
\and  Astronomical Observatory, Warsaw University, Al Ujazdowskie 4, 00478 Poland 
\and  New Mexico State University, Department of Astronomy, 1320 Frenger Mall, Las Cruces, NM 88003
\and Tombaugh Fellow}

\date{Received 00 month 0000 / Accepted 00 month 0000}

\abstract{}{We analyze the properties of VIRGO detector with the aim of studying
its ability to search for coalescing black hole binaries. We focus on
the remnants of the Population III stars, which currently should be massive
black holes ($\sim 100-1000~M_\odot$), some of them bound in binary
systems.  The coalescence of such binaries due to emission 
of gravitational waves may be currently observable.}
{We use a binary population synthesis to model the evolution of
Population III binaries.}
{
 We calculate the signal to noise ratios of gravitational waves emitted
by the system in each of the coalescence  phase: inspiral, merger and ringdown,
and provide simple formulae for the signal to noise ratio as a function of 
masses of the binaries.
We estimate the detection rates for the VIRGO interferometer and 
 also compare them  with the 
 estimates  for the current  LIGO. We show that 
these expected rates are similar to, or larger than the expected rates from 
coalescences of Population I and II compact object binaries.
}{}
\authorrunning{Kulczycki et al.}
\titlerunning{VIRGO sensitivity...}
\maketitle

\section{Introduction}

The high frequency gravitational wave detectors are 
approaching their specified sensitivities.  The LIGO 
observatory \citep{1992Sci...256..325A}
is already taking data, while  VIRGO  \citep{brada1990}
undergoes the engineering runs. It is therefore 
very important to study in detail the detectability 
of different types of gravitational wave sources with the current detectors.

A new class of sources - black hole binaries originating 
in population III stars - was recently suggested \citep{2004ApJ...608L..45B}.
\citet{2001ApJ...550..890B}   have shown that zero metallicity star are stable, and can form 
intermediate mass $\approx 100 - 500\,M_\odot$ black holes \citep{2002ApJ...567..532H}. 
They have shown that the stars 
with the initial masses between $140$ and $260\,M_\odot$
undergo violent pair instability explosions and leave no remnant 
while the stars outside these mass range leave black holes with essentially the
same mass as the initial star.
Numerical investigation of \citet{1999ApJ...527L...5B,2002ApJ...564...23B}  of Population III star formation indicate
that their initial mass function allowed a much 
larger number of massive stars than that of Population I.
\citet{2004ApJ...612..597W} have a  considered detectability 
of  gravitational 
waves emitted in the process of black hole growth
on the  Population III black hole seeds.

It is therefore important to ask a question whether 
Population III stars could form binaries. 
However,
it was recently shown by \citet{2004ApJ...615L..65S}
that a rotating metal free cloud may lead to formation 
of a metal free binary. Moreover, massive stars may form 
from clouds of slightly enhanced metallicity which may
cool off more efficiently. All known stellar populations  
contain a large fraction of binaries.

The  detectability analysis of Population III
black hole binaries was performed by
\citet{2004ApJ...608L..45B} for the case of the advanced LIGO detector. 
In that paper   
a simple model of the evolution of the Population III stars was developed.
The model was based on
numerical calculation of stellar evolution. 
A top heavy initial mass function was assumed, and 
a binary fraction of Population III stars of  10\%.
The number of stars was estimated with the assumption
that a fraction of $10^{-3}$ of the baryonic mass in the Universe was
processed in Population III binaries.

In this paper we investigate the detectability of 
the Population III black hole binaries with the 
current configuration of LIGO, and  with VIRGO at the expected sensitivity.
The signal to noise from a compact object 
coalescence for the LIGO detector has been calculated 
by \citet{1998PhRvD..57.4535F}. In section 2
 we estimate the signal to noise for VIRGO using 
the same formalism. The detection rate is calculated in section 3, 
while section 4 contains  conclusions and discussion.

\section{Signal to Noise Ratio}

To calculate the signal to noise ratio of gravitational wave from binary black
hole coalescence 
by the  VIRGO detector we  follow the formalism of \citet{1998PhRvD..57.4535F}. We find  the average of the squared signal to noise ratio (SNR) over all
orientations of and directions to the source, $< \rho^2 >$:
\begin{equation}
\label{rho2}
\bigg(\frac{S}{N}\bigg)^2 = < \rho^2 > = \frac{2(1+z)^2}{5\pi^2D(z)^2}
\int\limits_0^\infty \frac{1}{f^2S_\mathrm{h}(f)}
\frac{\mathrm{d}E_\mathrm{e}}{\mathrm{d}f_\mathrm{e}}[(1+z)f]~\mathrm{d}f\, ,
\end{equation}
where $z$ is the  redshift of the system,  $D(z)$ is the 
 luminosity distance,   %(which is defined by eq. \ref{lumdist}), 
$S_\mathrm{h}(f)$ is the 
one sided power spectral density of
the interferometer  and $\mathrm{d}E_\mathrm{e}/\mathrm{d}f_\mathrm{e}$ is the 
locally measured wave energy spectrum.
{ We also introduce the frequency $f$ at the source and
$f_\mathrm{e}=(1+z)^{-1}f$ at the interferometer.}
In this section we follow the system of units 
in  which $G=c=1$.

\begin{figure}
 \resizebox{\hsize}{!}{\includegraphics{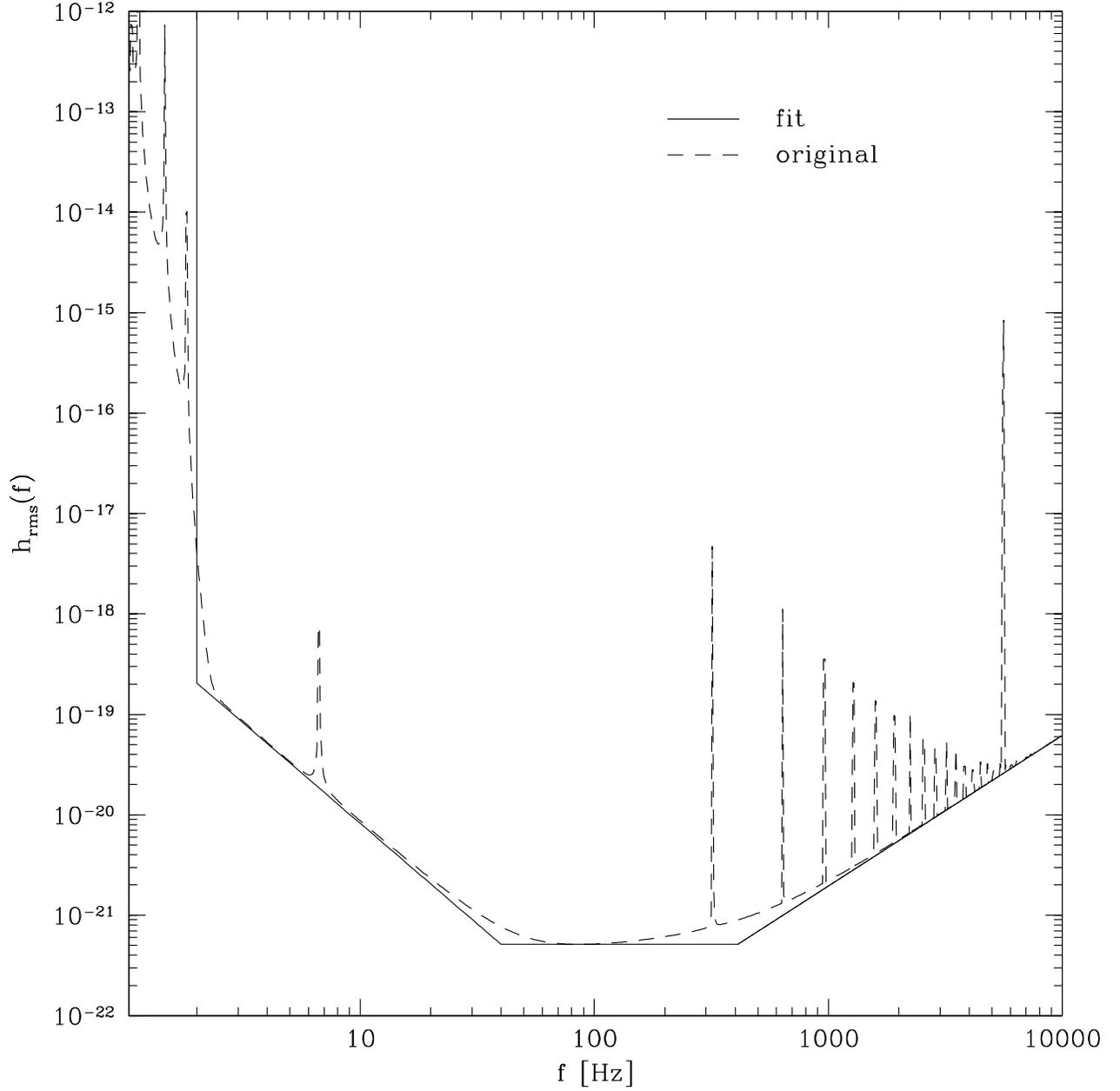}}
 \caption{The VIRGO sensitivity curve expressed as
$h_\mathrm{rms}(f)=\sqrt{f S_\mathrm{h}(f)}$ (dashed lines) and the approximation used in this work
 (solid
lines).}
 \label{fig_senscurve}
\end{figure}

\citet{1998PhRvD..57.4535F} approximated the sensitivity curve
of the detector { with the quantity  $h_\mathrm{rms}(f)=
\sqrt{fS_\mathrm{h}(f)}$, i.e.} root mean squared fluctuation in the noise at
frequency $f$ in a bandwidth $\Delta f=f$, and then fitting them
with  power laws. We have analogously approximated the VIRGO sensitivity
\citep{Virgo-noise} with a simple  analytical model:
\begin{eqnarray}
\label{hrms}
 h_\mathrm{rms}(f) & = & \left\{
    \begin{array}{llll}
      \infty & \mathrm{for}& f < f_\mathrm{s} \\
      h_\mathrm{m} (\frac{\alpha f}{f_\mathrm{m}})^{-2} & \mathrm{for} &
         f_\mathrm{s} \le f < \frac{f_\mathrm{m}}{\alpha} \\
      h_\mathrm{m} &  \mathrm{for} & \frac{f_\mathrm{m}}{\alpha} \le f \le \alpha f_\mathrm{m}\\
      h_\mathrm{m} (\frac{f}{\alpha f_\mathrm{m}})^{3/2} & \mathrm{for} &
         \alpha f_\mathrm{m} < f \\
    \end{array}\right.
\end{eqnarray}
with parameters:
\begin{eqnarray}
\label{fitparameters}
\left\{
    \begin{array}{llll}
      f_\mathrm{s}=2\ \mathrm{Hz} \\
      f_\mathrm{m}=128\ \mathrm{Hz} \\
      \alpha=3.2 \\
      h_\mathrm{m}=5.12\cdot10^{-22} \\
    \end{array}\right.
\end{eqnarray}
The VIRGO design  sensitivity curve and the fit are shown  in Figure~\ref{fig_senscurve}.
Our approximation  of the VIRGO sensitivity curve  differs from the 
 one  used for the initial
LIGO detector by \citet{1998PhRvD..57.4535F} in the low frequency domain. 
In this domain initial LIGO sensitivity
curve has been  approximated by a function $h_\mathrm{rms}(f)\sim f^{-3/2}$,
while the slope of the VIRGO sensitivity curve is 
{ steeper, with}  $h_\mathrm{rms}(f)\sim f^{-2}$, yet it extends further to the low
 frequency regime.
This modifies the analytical 
formulae  for the signal to noise ratio, especially for the 
high mass merging compact objects.

The energy spectrum $\mathrm{d}E_\mathrm{e}[(1+z)f]/\mathrm{d}f_\mathrm{e}$ has
different form depending on the coalescence phase: inspiral, merger or ringdown.
In the following  we have the same definitions of the energy spectra in these phases as  used by \citet{1998PhRvD..57.4535F}. For the inspiral phase 
we have used the  formula:
\begin{equation}
\label{dEdf_insp}
\frac{\mathrm{d}E}{\mathrm{d}f}=\frac{1}{3}\pi^{2/3}\mu M^{2/3}f^{-1/3}\, ,
\end{equation}
where $\mu$ is the reduced mass of the
system, while $M$ is its total mass,
  and $f<f_\mathrm{merge}=0.02/M$, the frequency at which the inspiral phase ends.
The merger phase energy spectrum was approximated 
 by a flat spectrum between $f_\mathrm{merge}$ and 
the
quasi normal mode frequency $f_\mathrm{qnr}=0.13/M$:
\begin{equation}
\label{dEdf_merg}
\frac{\mathrm{d}E}{\mathrm{d}f}=\frac{\epsilon_\mathrm{m} M F(\mu/M)}
{f_\mathrm{qnr}-f_\mathrm{merge}}~\Theta(f-f_\mathrm{merge})
~\Theta(f_\mathrm{qnr}-f),
\end{equation}
where $\epsilon_m$ is the fraction of mass energy radiated during merger phase
which we assumed to be $\epsilon_m=0.1$, $\Theta(x)$ is the step function,
and $F(\mu/M)=(4\mu/M)^2$.  Finally, the energy spectrum 
in the ringdown phase is 
\begin{eqnarray}
\nonumber
\frac{\mathrm{d}E}{\mathrm{d}f} & = & \frac{A^2M^2f^2}{32\pi^3\tau^2}
\bigg\{\frac{1}{[(f-f_\mathrm{qnr})^2+(2\pi\tau)^{-2}]^2}+ \\
\label{dEdf_ring}
& & +\frac{1}{[(f+f_\mathrm{qnr})^2+(2\pi\tau)^{-2}]^2}\bigg\} \approx \\
\label{delta}
& \approx & \frac{1}{8}A^2QM^2\delta(f-f_\mathrm{qnr})[1+\mathcal{O}(1/Q)]\, ,
\end{eqnarray}
where $Q=\pi\tau f_\mathrm{qnr}$ is the quality factor, 
$\tau$ is the  damping
time of the quasi-normal mode of the newly formed black hole,
and $A$ is the initial amplitude of the wave. The quality factor
is approximately $Q\approx 12$, we assume that the initial 
amplitude $A=0.4$. The amplitude scales as
$F(\mu/M)$ for non equal mass binaries.

\begin{figure}
 %\resizebox{\hsize}{!}
\begin{center}
{\includegraphics[width=0.95\columnwidth]{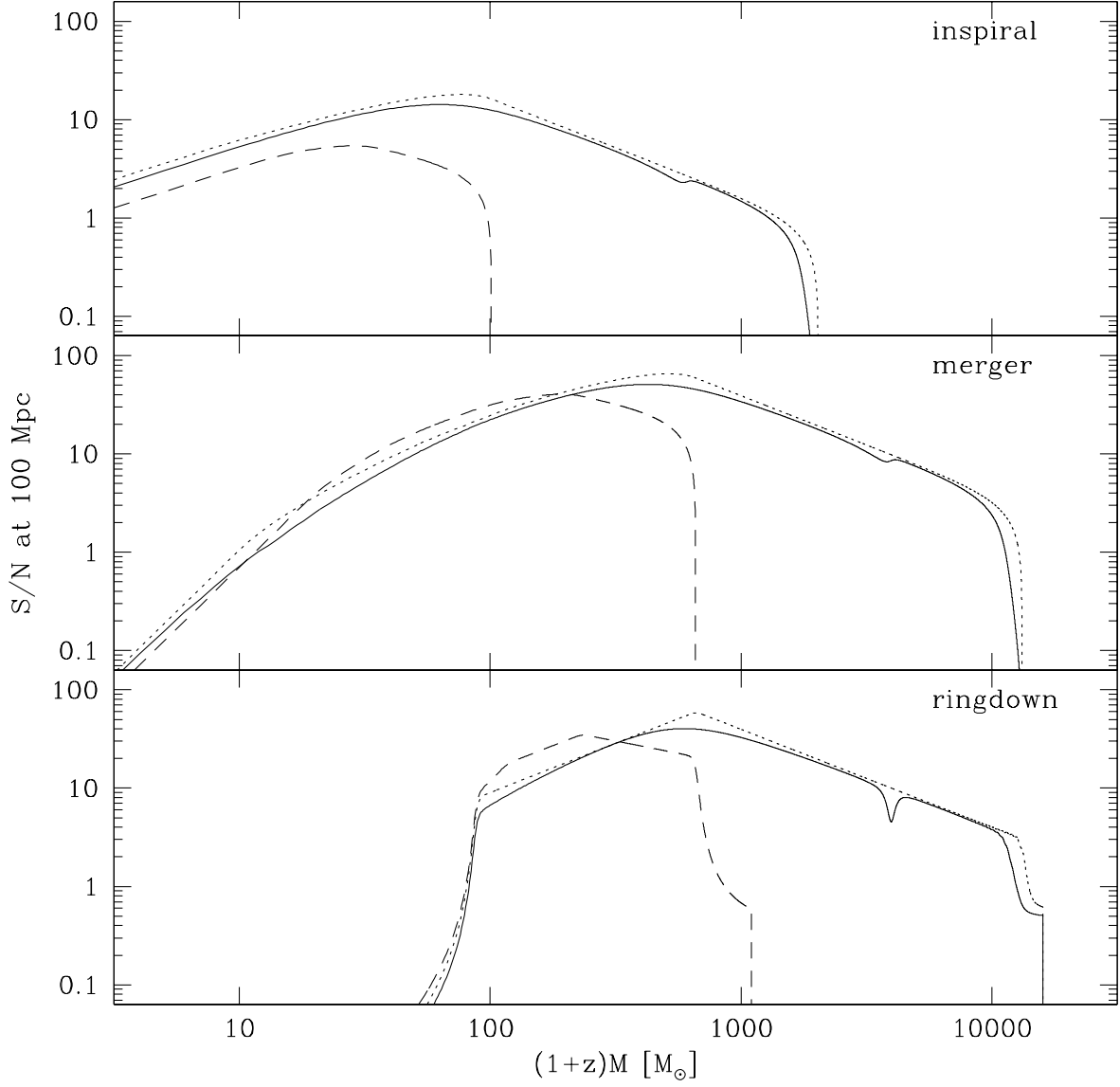}}
\end{center}
 \caption{SNR as a function of the redshifted total mass of a merging binary 
for VIRGO (solid lines) and initial LIGO (dashed lines). 
The dotted lines present  the  numerical results obtained using the 
approximate noise curve.
The upper panel shows SNR in the inspiral phase, the middle panel  
in the merger phase, and the lower panel in the ringdown phase.}
 \label{fig_snr}
\end{figure}

The details of the calculation are shown in the Appendix. 
Here we present the final results in a form useful for future calculations.
In the inspiral phase the signal to noise ratio is:
\begin{eqnarray}
\label{inspnum}
{S\over N} = {M_z^{5/6}\over D_{100}}\sqrt{4\mu \over M}\times\left\{
\begin{array}{lll}
 0.95 
\sqrt{1-3.26\times 10^{-5} {M_z}^{10/3} }
&\mathrm{for} & M_z<9.9  \\
1.21 \sqrt{1-0.197 M_z^{1/3} }
&\mathrm{for} &  9.9 <M_z< 102 \\
5.64\times 10^{-4} \sqrt{8.69\times 10^{12} M_z^{-11/3}-1}
&\mathrm{for} & 102<M_z<2027 \\
0 &\mathrm{for} & 2027 < M_z
\end{array} \right.
\end{eqnarray}
where $M_z = M(1+z)/M_\odot$, and $D_{100}=D/100$\,Mpc.
In the merger phase the signal to noise ratio is:

\begin{eqnarray}
\label{mergnum}
{S\over N} =  \displaystyle {1\over D_{100}} {4\mu \over M} \left\{
\begin{array}{lll}
\displaystyle 3.38\times 10^{-3} M_z^{5/2}                                              & \mathrm{for} & M_z<9.9\\
\displaystyle 9.13\times 10^{-2} M_z\sqrt{4\log\left(\displaystyle{M_z\over 9.9}\right)- {M_z^3\over 2.0\times 10^5}+{4\over 3} }& \mathrm{for} & 9.9<M_z<64.3\\
\displaystyle 0.25\times M_z & \mathrm{for} & 64.3<M_z<102\\
\displaystyle  9.13\times 10^{-2} M_z\sqrt{1+4\log\left(\displaystyle{659\over M_z}\right)-{1.05\times 10^8\over M_z^{4}}} & \mathrm{for} & 102<M_z<663\\
\displaystyle {3.96\times 10^4}{1 \over M_z} & \mathrm{for} & 663<M_z<2027\\
\displaystyle 9.13\times 10^{-2}  M_z\sqrt{{1.88\times 10^{11}\over M_z^{4}}-6.25\times 10^{-6}} & \mathrm{for} &2027<M_z<13169\\
0  & \mathrm{for} & 13169<M_z .
\end{array}\right.
\end{eqnarray}
Finally in the ringdown phase  the 
signal to noise is:
\begin{eqnarray}
\label{ringnum}
{S\over N} = \displaystyle {1\over D_{100}} {4\mu \over M} \left\{
\begin{array}{lll}
1.80\times 10^{-4} M_z^{5/2}  & \mathrm{for} & M_z<63.4\\
0.0938 M_z & \mathrm{for} & 63.4<M_z<650 \\
4.126\times 10^4 {1\over M_z} & \mathrm{for} & 650<M_z<13265 \\
0 & \mathrm{for} & 13265<M_z .
\end{array}\right. 
\end{eqnarray}

We have calculated the SNR using the above approximate formulae, as well
as numerically integrating the noise spectrum   given in \citep{Virgo-noise}.
The results are shown in Figure~\ref{fig_snr}. We also show for comparison the 
SNR for the initial LIGO configuration. 
The inspiral phase signal to noise decreases quickly with increasing mass
when $(1+z)M> 100 M_\odot$ for VIRGO.
However, for VIRGO the merger and  ringdown phase S/N peaks at about 
$800-1000 M_\odot$, where the LIGO sensitivity already drops down.
Thus  VIRGO  may be able to catch the signal of 
black hole binary coalescence with the total masses above 500\,M$_\odot$.
The sensitivity 
of  VIRGO detector is similar to that of the initial LIGO for the merging binaries
 with the masses below $200\,M_\odot$.
However, the difference in sensitivity in the low frequency regime
leads to improvement of the expected S/N ratio in VIRGO
for large redshifted masses.

\section{Detection Rate}

In order to estimate the detection rate in VIRGO and LIGO we 
simulate the evolution of massive Population III stars, using 
the  models of \citet{2004ApJ...608L..45B}. 
We made the  same assumptions about the
evolution of Population III stars: the initial mass function 
 is assumed to have a slope of $-2$ 
above $100\,M_\odot$ and extends up to $500\,M_\odot$.
We use the numerical results of \citet{2001ApJ...550..890B,2001A&A...371..152M,2002ApJ...567..532H} to 
describe the stellar evolution tracks, and calculate the He core masses with the
results of \citet{2002ApJ...567..532H}. The outcome 
of the nuclear evolution follows from \citet{2003ApJ...591..288H}.
We include pair instability supernovae so stars 
with initial masses between $140$ and $260 M_\odot$ leave no remnants.
Similarly to { the} population synthesis of {Population I stars,} the 
initial orbits are drawn from a distribution flat in logarithm, and the initial 
mass ratio distribution is flat. 
The binary fraction is assumed to be 10\%.
For more details see \citet{2004ApJ...608L..45B}.
This model of evolution of Population III binaries provides
an estimate of the distribution of masses and lifetimes of the
binaries.
We assume a  constant star formation rate between redshift
10 and 30, and obtain  the black hole  binary birth distribution function   $F(M,t,z) = \Theta(z-10)\Theta(30-z)\sum\delta(M-M_i)\delta(t-t_i)$,
where the sum goes over all the systems in the simulation.
About $10^{-3}$ of the baryonic mass of the Universe must have been
processed through Population III stars \citep{2001ApJ...551L..27M}
which leads to  the comoving rate
of Population III star formation of \mbox{$R_\mathrm{sfr}\approx1.4\cdot10^{-2}
M_\odot \mathrm{Mpc}^{-3}\mathrm{yr}^{-1}$}.
 { The calculations are carried out for a flat cosmological model with
 $\Omega_\mathrm{m}=0.3$, $\Omega_\Lambda=0.7$ and
the Hubble constant $H_0=67~\mathrm{km}/
\mathrm{s}/\mathrm{Mpc}$.}
The luminosity distance, $D(z)$, is given by
 (see e.g. \citet{2004A&A...415..407B}) 
\begin{equation}
\label{lumdist}
D(z)= c \, (1+z)\int\limits_\mathrm{0}^z (1+z')\bigg|\frac{\mathrm{d}t}{\mathrm{d}z'}
    \bigg|~\mathrm{d}z',
\end{equation}
where $t$ is the cosmic time:
\begin{equation}
\bigg|\frac{\mathrm{d}t}{\mathrm{d}z}\bigg|=\frac{1}{H_0(1+z)
  \sqrt{\Omega_\mathrm{m}(1+z)^3+\Omega_\Lambda}}\, .
\end{equation}

\begin{figure*}
\includegraphics[width=0.47\columnwidth]{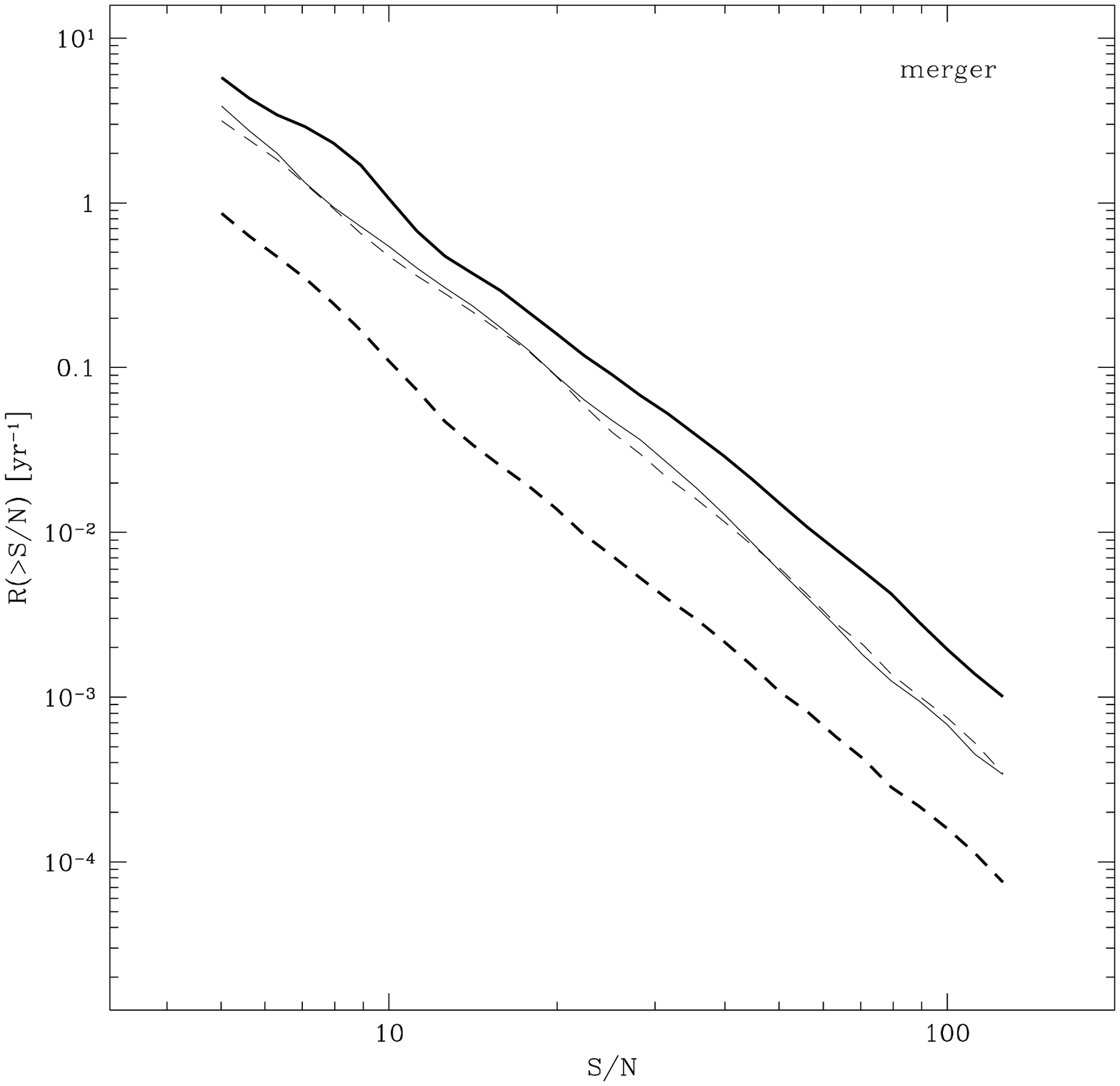}
\includegraphics[width=0.47\columnwidth]{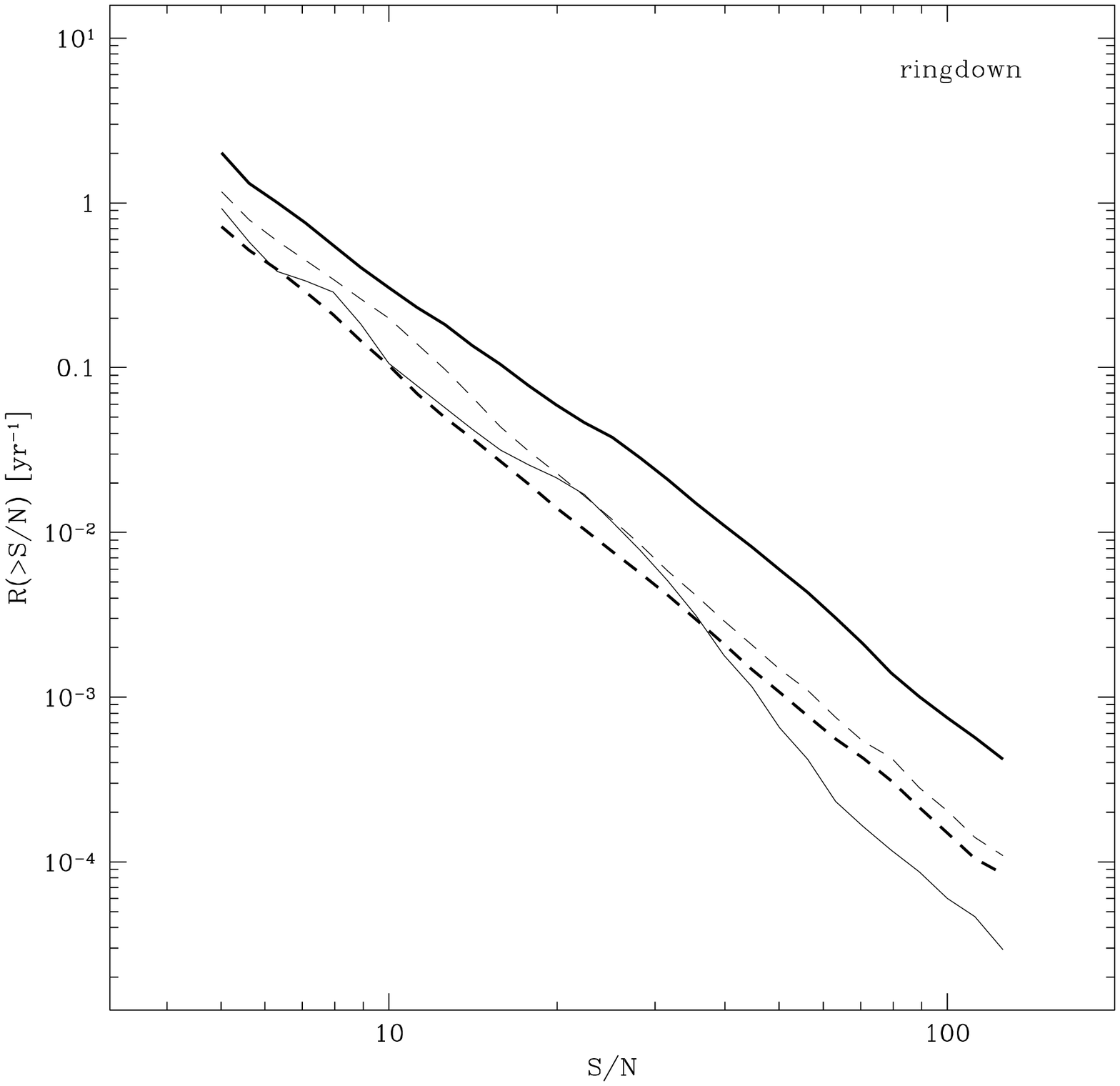}

 \caption{Expected observed coalescence rate of Population III black hole
binaries as a function of SNR for the VIRGO (solid lines) and initial LIGO
(short dashed lines) detectors, while the long dashed lines 
show the expected rates in the advanced LIGO \citep{2004ApJ...608L..45B}.
 The thick lines correspond to the case withour hardening of the binaries
while the thin lines show the case when all binaries are hardened by 
a factor of
$\sim10$. Left  panel corresponds to the  merger phase, 
and the right panel to the
 ringdown
phase.}
 \label{fig_R}
\end{figure*}

\begin{figure}
 \resizebox{\hsize}{!}{\includegraphics{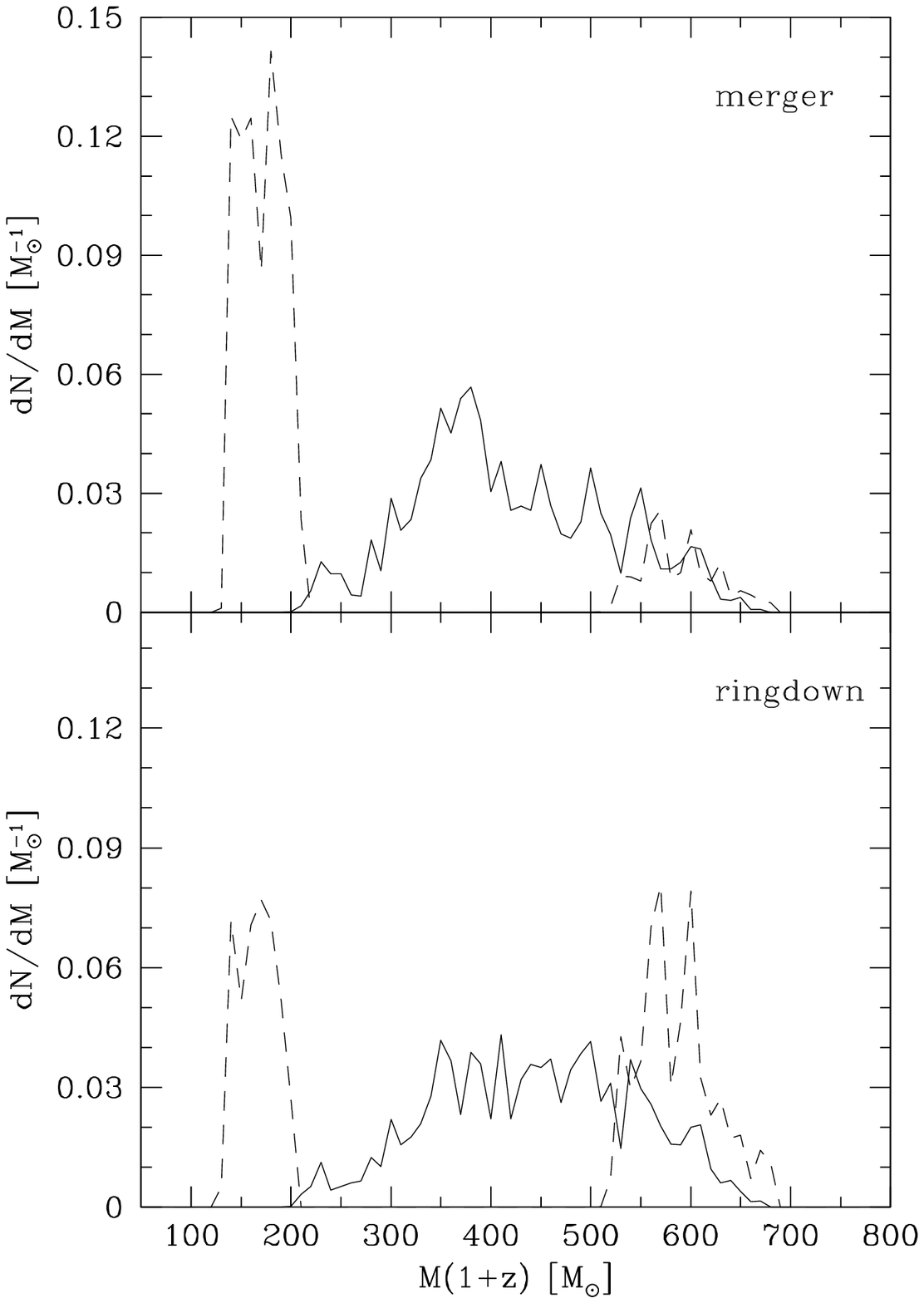}}
 \caption{Differential rate as a function of observed redshifted mass of black hole
binary system.  The solid lines correspond to the case with no hardening 
and the dashed lines represent the case with hardening by a factor of $\sim10$.
 The upper panel presents the case of detection in the 
 merger phase, while the lower panel shows the case of the  ringdown phase detection.
While the binaries were formed at
 redshifts from $10$ to $30$, the detectable mergers happen at redshifts below unity.}
 \label{fig_dRdM}
\end{figure}

The  observed  black hole binaries coalescence rate  is given by:
\begin{equation}
\frac{\mathrm{d}R}{\mathrm{d}M}=\int\limits_0^{z_\mathrm{max}}
   \frac{\mathrm{d}f_\mathrm{coal}(z)}{\mathrm{d}M}\frac{1}{1+z}
   \frac{\mathrm{d}V}{\mathrm{d}z}~\mathrm{d}z, 
\end{equation}
where he maximum redshift $z_\mathrm{max}$ is estimated using equations (\ref{sn_insp}), (\ref{sn_merg}) and
(\ref{sn_ring})
for the three phases of coalescence respectively, for a given  SNR value in the detector.
 The comoving volume element is:
\begin{displaymath}
\label{dVdz}
\frac{\mathrm{d}V}{\mathrm{d}z}=\frac{4\pi c  D(z)^2}{1+z}\bigg|
   \frac{\mathrm{d}t}{\mathrm{d}z}\bigg|.
\end{displaymath}
The black hole binary  coalescence rate, $\mathrm{d}f_\mathrm{coal}(z)/\mathrm{d}M$,
depends on the birth rate  of the binaries retarded by their 
evolutionary time:
\begin{equation}
\frac{\mathrm{d}f_\mathrm{coal}(z)}{\mathrm{d}M}=\int~\mathrm{d}t'
    ~F(M,t',z_\mathrm{f}),
\end{equation}
where $z_\mathrm{f}$ is the source formation redshift, which
 can be found by solving the
following integral:
\begin{equation}
t'=\int\limits_{z}^{z_\mathrm{f}}~\bigg|
   \frac{\mathrm{d}t}{\mathrm{d}z}\bigg|~\mathrm{d}z.
\end{equation}
In the calculation of the binary inspiral time we also consider the possibility
that there are additional factors leading to tightening
of the orbits. These may be due to interaction with the ISM or
with stars in dense centers of Galaxies, where such binaries are likely to sink.
We model these processes by considering a model in which 
the orbits are additionally tightened (the binaries become harder)
 by a factor of ten.

We present the results of the rate calculation as a function of the signal to noise
in Figure~\ref{fig_R}. Each  plot contains two sets of curves:
the rate for the VIRGO detector, the rate for the initial LIGO.
These rates are lower by a factor 
about $\sim 10^4$ than   the  ones calculated by \citet{2004ApJ...608L..45B}.
In each   case  we present the rates assuming that the binaries
evolve without interacting with other stars or gas, and with the assumption
that they are hardened by a factor of 10.

The detection rates do not depend strongly on the assumed evolution of binaries
and are similar in the merger and the inspiral phase.
 For the assumed $S/N\approx 10$ the detection rate without hardening is $R(>S/N)
\approx (1-10)~\mathrm{yr}^{-1}$. It drops with increasing required SNR and for
$S/N\approx20$ it is about ten times less ($R(>S/N)\approx (0.1-1)
~\mathrm{yr}^{-1}$).  Since the required SNR in the detector 
corresponds roughly to the maximal distance out to which coalescences are detectable 
the rate drops with as $(S/N)^{-3}$. On the other hand the rate increases with
the inverse cube of the detector sensitivity - the minimum detectable strain amplitude.

In Figure~\ref{fig_dRdM} we present the 
distribution of total masses of the binaries 
detectable in merger and ringdown phases for the cases 
with and without hardening in the VIRGO detector. 
The distributions are normalized to unity.
These distributions show the most probable masses 
of the coalescing binaries in the framework of the given model of 
evolution of black hole binaries:  without hardening and with hardening by a factor of 10. 
The shape of the distribution appears to depend strongly on the
particular choice of the hardening. Therefore
the optimal search strategy would be to search the entire 
spectrum of the masses from $100\,M_\odot$ to $1000\,M_\odot$, 
as the current models 
can not provide more accurate predictions. The best 
strategy is to look for bursts of single frequency 
signals from the ringdown phase.

\section{Summary}

We have calculated the signal to noise ratio from 
coalescing black hole binaries in VIRGO.
The low frequency sensitivity of this detector
leads to increased sensitivity to 
intermediate mass black hole mergers.
We find that for the merger and ringdown signals
VIRGO has a maximum of sensitivity for systems\
with the redshifted 
masses in the range from $500\,M_\odot$ and $2000\,M_\odot$.
This makes VIRGO an especially valuable 
instrument in search for the signals from the intermediate mass
black hole coalescences, as well as captures of stars
by intermediate mass black holes in clusters.

Population  III stars must have included a large number of high mass
stars which should have left a significant number of black holes
with masses above $100\,M_\odot$. 
Some of them could have been formed binaries.
Based on the 
evolutionary scenarios  of such stars we calculate the properties
of such binaries and estimate the rate with which they 
should be detectable by VIRGO. 
 The rate is of the order of 
one detection per year for the design sensitivity of initial VIRGO
and similar for the current LIGO sensitivity.
While this number  is not large by itself 
we note that it is comparable or even larger
than the  
expected detection rate of the 
most often considered  gravitational wave sources i.e.
 the double neutron star binaries \citep{2002ApJ...572..407B,2004ApJ...601L.179K}.
It should be noted that the number of the
detectable colaescences of black hole binaries originating 
in Population III stars increases quickly with the detector sensitivity.
For the Advanced LIGO the expected rates are larger 
by a factor of $\sim 10^4$ \citep{2004ApJ...608L..45B}.
Therefore even modest improvements of the sensitivity 
will make the detection of these binaries more likely.

One should be still cautious when dealing with 
estimates of the properties of the Population III stars.
The rate we calculate depends on the actual shape 
of the initial mass function of Population III stars, and 
on their  binar fraction. The knowledge of evolution of Population III 
stars is only known from numerical models, and hence the models
evolution of population III binaries is only approximate.
However, these uncertainties and the possible role that Population III
star could have played in the early Universe makes it even more interesting to 
look for traces of their existence, and gravitational waves searches provide such an
opportunity.

Therefore we consider that it is very important to 
 include the ringdown and merger 
templates in the search for merging binaries, and to look
for signals from merging binaries 
across the entire sensitivity range of the interferometric gravitational wave detectors.

\acknowledgements{This work was supported by the 
KBN grant 1P03D002228.
}

\bibliographystyle{aa}
%\bibliography{tekst}

\newpage

\appendix
\section{General formulae for the signal to noise ratio}

In order to find the general formulae for the signal to noise ratio
we 
insert equations (\ref{dEdf_insp}), (\ref{dEdf_merg}), (\ref{dEdf_ring}) and (\ref{hrms}) 
into equation
(\ref{rho2}) and perform the integration.
We introduce a dimesionless variable
\begin{equation}
v=\frac{(1+z)~\alpha f_\mathrm{m}}{f_\mathrm{merge}}.
\end{equation}
For the inspiral phase we obtain:
\begin{eqnarray}
\label{sn_insp}
\frac{S}{N} = \left\{
 \begin{array}{lll}
   \sqrt{{\cal F}_\mathrm{i}\left[12\alpha^{1/3}-\frac{99}{10}\alpha^{-1/3}
      -\frac{11}{10}\alpha^{-1/3}v^{10/3}-
      \alpha^4(\frac{f_\mathrm{s}}{f_\mathrm{m}})^{11/3}\right]}& {\rm for\ \ }& v<1\\
   \sqrt{{\cal F}_\mathrm{i}\left[12\alpha^{1/3}-11(\frac{v}{\alpha})^{1/3}-
      \alpha^4(\frac{f_\mathrm{s}}{f_\mathrm{m}})^{11/3}\right]} &{\rm for\ \ } & 1<v<\alpha^2\\
   \sqrt{{\cal F}_\mathrm{i}\left[\alpha^{1/3}(\frac{\alpha^2}{v})^{11/3}-
      \alpha^4(\frac{f_\mathrm{s}}{f_\mathrm{m}})^{11/3}\right]} &{\rm for\ \ }&\alpha^2<v<\alpha f_m /f_s\\
   0& {\rm for\ \ } & \alpha f_m/f_s < v \\
 \end{array}\right. 
\end{eqnarray}
where
\begin{displaymath}
{\cal F}_\mathrm{i}=\frac{2[(1+z)M]^{5/3}[4\mu/M]}{55\pi^{4/3}D(z)^2
h_\mathrm{m}^2f_\mathrm{m}^{1/3}}.
\end{displaymath}

For the merger phase we obtain:
\begin{eqnarray}
\label{sn_merg}
\frac{S}{N} = \left\{
 \begin{array}{lll}
   \sqrt{{\cal F}_\mathrm{m}[\frac{4}{3}v^3
      \frac{\kappa^3-1}{\kappa^3}]} &{\rm for\ \ } & v<1\\
   \sqrt{{\cal F}_\mathrm{m}[4\ln v-\frac{4}{3}
      (\frac{v^3-\kappa^3}{\kappa^3})]} & {\rm for\ \ } &
     1<v<\kappa\\
\sqrt{{\cal F}_\mathrm{m}[4\ln \kappa]} &{\rm for\ \ } &
\kappa< v<\alpha^2 \\
   \sqrt{{\cal F}_\mathrm{m}[1+4\ln
       (\frac{\kappa\alpha^2}{v})-\frac{\alpha^8}{v^4}]} &
       {\rm for\ \ } & \alpha^2<v<\kappa\alpha^2 \\
   \sqrt{{\cal F}_\mathrm{m}[\alpha^8v^{-4}
       (\kappa^4-1)]} &{\rm for\ \ }&\kappa\alpha^2<v<\alpha f_m/f_s \\
   \sqrt{{\cal F}_\mathrm{m}[(
       \frac{\kappa\alpha^2}{v})^4-(\frac{\alpha f_\mathrm{s}}
       {f_\mathrm{m}})^4]} &{\rm for\ \ }& \alpha f_m/f_s<v<\kappa\alpha f_m/f_s\\
   0&{\rm for\ \ }&\kappa\alpha f_m/f_s<v\\
 \end{array}\right.
\end{eqnarray}
where
\begin{displaymath}
\kappa=f_\mathrm{qnr}/f_\mathrm{merge}
\end{displaymath}
and
\begin{displaymath}
{\cal F}_\mathrm{m}=\frac{\epsilon_\mathrm{m}
M(1+z)^2(4\mu/M)^2}{10\pi^2 D(z)^2h_m^2f_\mathrm{merge}(\kappa-1)}.
\end{displaymath}
The fraction  of the mass emitted in gravitational waves 
during the merger phase has been assumed to be $\epsilon_\mathrm{m}=0.1$
\citep{1998PhRvD..57.4535F}.

For the ringdown phase the result is
\begin{equation}
\label{sn_ring}
\frac{S}{N}=\sqrt{\frac{(1+z)^3M^2A^2Q(4\mu/M)^2}
{20\pi^2D(z)^2f_\mathrm{qnr}~S_h[f_\mathrm{qnr}/(1+z)]}}.
\end{equation}
Note that the signal to noise in the merger phase scales as $\epsilon_\mathrm{m}^{1/2}$,
and in the ringdown phase it scales linearly with the initial amplitude $A$.

\end{document}